\documentclass[a4paper,preprint,eqsecnum,showpacs,prb]{revtex4}
\usepackage[T1]{fontenc}
\usepackage[latin1]{inputenc}
\usepackage{graphicx}
\usepackage{amssymb}

\makeatletter

\newcommand{\boldsymbol}[1]{\mbox{\boldmath $#1$}}

\makeatother
\begin{document}

\title{Analysis of optical magnetoelectric effect in GaFeO$_3$}

\author{Jun-ichi Igarashi$^1$ and Tatsuya Nagao$^2$}

\affiliation{$^1$Faculty of Science, Ibaraki Univ., Mito,
Ibaraki 310-8512, Japan \\
$^2$Faculty of Engineering, Gunma Univ., Kiryu, 
Gunma 376-8515, Japan}

\begin{abstract}

We study the optical absorption spectra in a polar ferrimagnet GaFeO$_3$. 
We consider the $E1$, $E2$ and $M1$ processes on Fe atoms.
It is shown that the magnetoelectric effect on the absorption spectra 
arises from the $E1$-$M1$ interference process through the hybridization 
between the $4p$ and $3d$ states in the noncentrosymmetry environment of 
Fe atoms. We perform a microscopic calculation of the spectra on a cluster 
model of FeO$_6$ consisting of an octahedron of O atoms and an Fe atom 
displaced from the center with reasonable values for Coulomb interaction and 
hybridization. 
We obtain the magnetoelectric spectra, which depend on the direction of 
magnetization, as a function of photon energy in the optical region 
$1.0-2.5$ eV, in agreement with the experiment.

\end{abstract}

\pacs{78.20.Ls, 78.20.Bh, 78.40.-q}

\maketitle

\section{Introduction}

It is known that the breaking of time-reversal symmetry in magnetic materials 
gives rise to interesting magneto-optical effects such as the double circular 
reflection for circularly polarized light and the Faraday effect for linearly 
polarized light.\cite{Landau}
When the spatial inversion symmetry is further broken, for example,
in polar or chiral  materials, novel magneto-optical effects were expected 
to come out.\cite{Hornreich1968}
Those effects are known as the nonreciprocal directional 
dichroism or magnetochiral dichroism, and have been extensively studied.
\cite{Markelov1977,Bubis1982,Rikken1997,Kleindienst1998,Rikken2002}
Among a variety of compounds, Cr$_2$O$_3$ is one of the most investigated 
compounds. The magnetoelectric effect, that is, a linear relation between 
the magnetic and electric fields in matter was proved in 1950s.
\cite{Dzyaloshinskii1959,Astrov1960} Later, the nonreciprocal rotation and 
ellipticity of light were measured,\cite{Krichevtsov1996} 
and were successfully analyzed by using a ligand field model for Cr atoms.
\cite{Muto1998}

Another notable compound is GaFeO$_3$, which was first synthesized by Remeika.
\cite{Remeika1960} This compound exhibits simultaneously spontaneous 
electric polarization and magnetization at low temperatures. 
The large magnetoelectric effect was observed by Rado.\cite{Rado1964}
Recently, untwinned large single crystals have been prepared,\cite{Arima2004} 
and the optical absorption measurement has been carried out 
with changing the direction of magnetization.\cite{Jung2004}
It has been found that the absorption intensity in the region of photon energy 
$1.0-2.5$ eV changes with reversing the direction of the magnetization.
The purpose of this paper is to analyze in detail this phenomenon 
by carrying out a microscopic calculation of the spectra 
and to elucidate the microscopic origin. 
Although several qualitative arguments have been done,\cite{Jung2004,Ogawa2004}
as far as we know, the spectra have not been calculated yet as a function 
of photon energy.

\begin{figure}[h]
\begin{center}\includegraphics[%
  scale=0.5]{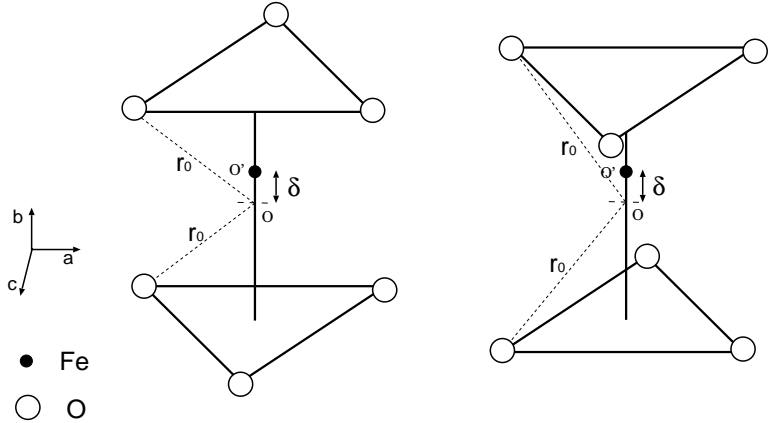}
\caption{Two kinds of octahedrons of oxygen atoms (white circles). 
Fe atoms (black circles) are displaced from the center of the octahedron O to 
the off-center O' along the $b$ axis by amount $\delta$; $\delta=0.26
\textrm{\AA}$ at
Fe1 sites and $\delta=-0.11 \textrm{\AA}$ at Fe2 sites.
\label{octahedron}}
\end{center}
\end{figure}

The crystal of GaFeO$_3$ has an orthorhombic unit cell with the space group
$Pc2_{1}n$.\cite{Wood1960}
The magnetic moments at Fe1 and Fe2 sites align antiferromagnetically 
along the $\pm c$ axis. The actual compound, however, behaves as a ferrimagnet,
\cite{Frankel1965} which reason is inferred that the Fe occupation at Fe1 
and Fe2 sites are slightly different from each other.\cite{Arima2004}
Each Fe atom is octahedrally surrounded by O atoms,
and slightly displaced from the center of the octahedron;
the shift is $0.26 \textrm{\AA}$ at Fe1 sites 
and $-0.11 \textrm{\AA}$ at Fe2 sites along the $b$ axis.
\cite{Arima2004} Thereby the spontaneous electric polarization is 
generated along the $b$ axis. We neglect slight distortion of octahedrons,
since their contributions are expected to be small to the $E1$-$M1$ terms.
There are two kinds of octahedrons with respect to the direction 
of Fe shift, as illustrated in Fig.~\ref{octahedron}.

\begin{figure}[h]
\begin{center}\includegraphics[%
  scale=0.5]{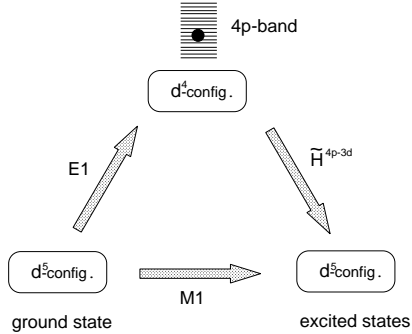}
\caption{
Illustration of the $E1$-$M1$ interference process. The black circle indicates
the presence of an electron in the $4p$ band.
\label{process}}
\end{center}
\end{figure}

In the analysis of optical absorption, we assume that the photon propagates 
along the $a$ axis in accordance with the experimental situation.
\cite{Jung2004} 
Restricting the processes only on Fe atoms, we derive the explicit forms of 
$E1$, $E2$, and $M1$ transitions.
We find that the $E2$ transition matrix elements are much smaller than those
of the $E1$ and $M1$ transitions. In addition to the $E1$-$E1$ and
$M1$-$M1$ processes, the $E1$-$M1$ interference process could have finite
contribution to the optical absorption through the mixing of 
the $3d^{4}4p$-configuration to the $3d^5$-configuration, as illustrated in 
Fig.~\ref{process}.
Such mixings are the result of the 
noncentrosymmetric environment on Fe atoms. 
In order to describe such processes,
we employ a cluster model of FeO$_{6}$, which includes all the $3d$ and $4p$
orbitals of Fe atoms and the 2p orbitals of O atoms.
The Coulomb interaction and the spin-orbit interaction are taken into account
in the $3d$ orbitals. Since Fe atoms are located in the noncentrosymmetric 
environment, the $4p$ and $3d$ states could be coupled to each other.
A similar cluster model has been considered 
in the analysis of resonant x-ray scattering in magnetite,
\cite{Igarashi2008}
where Fe atoms at A sites are in the noncentrosymmetric environment,
at the center of tetrahedrons of O atoms.
Deriving an effective hybridization between the $4p$ and $3d$ states 
as well as a ligand field on the $3d$ states through the hybridization 
with the O $2p$ states, we diagonalize the Hamiltonian matrix 
in the $3d^5$- and $3d^4$-configurations to obtain the energy eigenstates.
These states are used to calculate the absorption spectra.

In the experiment, the magnetic field was applied along the $\pm c$ axis, and
the difference of the absorption spectra between the two directions was 
measured, which would be termed as ``magnetoelectric" spectra.\cite{Jung2004} 
Since the compound is a ferrimanget, reversing direction of applied magnetic 
field results in reversing the direction of the local magnetic moment on Fe
atoms. Neglecting a small deviation from a perfect antiferromagnet,
we simply assume that the direction of the local magnetic moment is simply
reversed. We derive a formula for the magnetoelectric spectra which arise
from the $E1$-$M1$ process. Using this formula, we discuss various symmetry 
relations for the $E1$-$M1$ process and the relation to the nonreciprocal
directional dichroism and the anapole moment on these bases.
Finally, we carry out a microscopic calculation of the spectra arising 
from the $E1$-$M1$ process using the results of the FeO$_6$ cluster model.
We find the spectra as a function of photon energy in agreement with the 
experiment.\cite{Jung2004} 

This paper is organized as follows. In Sec. II, we introduce a cluster model
around Fe atoms. In Sec. III, we describe the optical transition operators
associated with Fe atoms.
In Sec. IV, we derive the formulas of the optical absorption, and
present the calculated spectra in comparison with the experiment.
The last section is devoted to concluding remarks.

\section{Electronic Structures around F\lowercase{e} atoms}

\subsection{Crystal electric field}

We start by examining the crystal electric field around the off-center 
position O' $=(0,0,\delta)$ to see the effect of lowering symmetry from
the cubic to trigonal ones.
Let charge $q$ be placed at the apexes of the octahedron. 
Then, the electrostatic potential $\phi(x,y,z)$ is expanded as
\begin{equation}
 \phi(x,y,z)= V_0 + \delta V_1 + \delta^2 V_2 + \cdots,
\label{eq.crys}
\end{equation}
with
\begin{eqnarray}
 V_0 &=& \frac{6q}{r_0}
          -\frac{7q}{24r_0^5}\bigl\{35z^4 -30z^2 r^2 + 3r^4 
          \pm 20\sqrt{2}z(x^3-3xy^2)\bigr\}, \\
 V_1 &=& -\frac{14q}{3r_0^5}\bigl\{5z^3-3zr^2\pm\frac{5}{4}\sqrt{2}
                            (x^3-3xy^2)\bigr\}, \\
 V_2 &=& -\frac{7q}{r_0^5}\bigl\{2z^2-(x^2+y^2) \bigr\},
\end{eqnarray}
where the $x$, $y$ and $z$ axes are along the crystal $c$, a, and b axes,
respectively, with the origin O'.
The distance between the center of the octahedron and the apexes 
is defined as $r_0$ and $r=\sqrt{x^2+y^2+z^2}$. 
The upper and lower signs correspond to the octahedron on the left and
right panels in Fig.~\ref{octahedron}, respectively. 
Term $V_0$ represents the so-called cubic field term,
which gives rise to a splitting of energy between $e_g$ and $t_{2g}$ states
in $3d$ orbitals. Term $V_1$ gives rise to a coupling between $3d$ and $4p$ 
states, and $V_2$ gives rise to additional splittings of energy within the $3d$ 
states as well as the $4p$ states. These forms are inferred to be correct
in symmetry point of view, but the covalency between Fe and O is, however,
expected to give rise to a similar but much larger effect.
We neglect the small point charge effect, and consider only the covalency
effect discussed in the following.

\subsection{Hamiltonian for a FeO$_6$ cluster}

We now introduce the Hamiltonian of a FeO$_6$ cluster, and derive the ligand
field on the $3d$ states and the effective hybridization between the $3d$ and 
$4p$ states. With the $2p$ states in O atoms in addition to the $3d$ and 
$4p$ states in the Fe atom,
we write the Hamiltonian as
\begin{equation}
 H = H^{3d} + H^{2p} + H_{\rm hyb}^{3d-2p} + H^{4p} 
 + H_{\rm hyb}^{4p-2p}, \label{eq.Ham}
\end{equation}
where 
\begin{eqnarray}
 H^{3d} & = & \sum_{m\sigma}E_m ^{d}d^{\dagger}_{m\sigma}d_{m\sigma}
  + \frac{1}{2}\sum_{\nu_{1}\nu_{2}\nu_{3}\nu_{4}}
  g\left(\nu_{1}\nu_{2};\nu_{3}\nu_{4}\right)d_{\nu_{1}}^{\dagger}
  d_{\nu_{2}}^{\dagger}d_{\nu_{4}}d_{\nu_{3}} \nonumber\\
& + & \zeta_{3d}\sum_{mm'\sigma\sigma'}
 \langle m\sigma|{\bf L}\cdot{\bf S}|m'\sigma'\rangle
  d^{\dagger}_{m\sigma}d_{m'\sigma'}.
 +{\bf H}_{\rm xc}\cdot \sum_{m\sigma\sigma'}
   ({\bf S})_{\sigma\sigma'}d^{\dagger}_{m\sigma}d_{m\sigma'},
\label{eq.H3d}\\
 H^{2p} &=& \sum_{j\eta\sigma}E^{p}p^{\dagger}_{j\eta\sigma}p_{j\eta\sigma},\\
 H_{\rm hyb}^{3d-2p} &=& \sum_{j\eta\sigma m}t_{m\eta}^{3d-2p}(j)
   d_{m\sigma}^{\dagger}p_{j\eta\sigma}+{\rm H.c.}, \\
 H^{4p} &=& \sum_{{\bf k}\eta'\sigma}\epsilon_{4p}({\bf k})
   p'^{\dagger}_{{\bf k}\eta'\sigma} p'_{{\bf k}\eta'\sigma},\\
 H_{\rm hyb}^{4p-2p} &=& \sum_{j\eta\sigma\eta'} 
 t_{\eta'\eta}^{4p-2p}(j)
 p'^{\dagger}_{\eta'\sigma} p_{j\eta\sigma}+{\rm H.c.}.
\end{eqnarray}
The $H^{3d}$ describes the energy of $3d$ electrons, where
$d_{m\sigma}$ represents an annihilation operator of a $3d$ electron 
with spin $\sigma$ and orbital $m$ ($=x^2-y^2,3z^2-r^2,yz,zx,xy$).
The second term in Eq.~(\ref{eq.H3d}) represents the 
intra-atomic Coulomb interaction with the matrix element 
$g\left(\nu_{1}\nu_{2};\nu_{3}\nu_{4}\right)$ expressed in terms of the
Slater integrals $F^{0}$, 
$F^{2}$, and $F^{4}$ ($\nu$ stands for $\left(m,\sigma\right)$).
The third term in Eq.~(\ref{eq.H3d}) represents the spin-orbit interaction 
for $3d$ electrons.
We evaluate atomic values of $F^2$, $F^4$, $\zeta_{3d}$ 
within the Hartree-Fock (HF) approximation,\cite{Cowan1981} and multiply 
$0.8$ to these atomic values in order to take account of the slight
screening effect. On the other hand, we  multiply 0.25 to the atomic value 
for $F^0$, since $F^{0}$ is known to be considerably screened 
by solid-state effects.
The last term in Eq.~(\ref{eq.H3d}) describes the energy arising from 
the exchange interaction with neighboring Fe atoms,
where $({\bf S})_{\sigma\sigma'}$ represents the matrix element of
the spin operator of $3d$ electrons. 
The exchange field ${\bf H}_{\rm xc}$ here has a dimension of energy,
and is $\sim k_{\rm B}T_{c}/4$ with $T_{c}\sim 250$ K. Note that this term is
served as selecting the ground state by lifting the degeneracy and therefore
the spectra depend little on its absolute value. The ${\bf H}_{\rm xc}$ 
is directed to the negative direction of the $c$ axis at Fe1 sites 
when the external magnetic field is applied along the positive direction
of the $c$ axis.

The $H^{2p}$ represents the energy of oxygen $2p$ electrons, where 
$p_{j\eta\sigma}$ is the annihilation operator of the $2p$ state with
$\eta=x,y,z$ and spin $\sigma$ at the oxygen site $j$.
The Coulomb interaction is neglected in oxygen 2p states. 
The $H^{3d-2p}_{\rm hyb}$ denotes the hybridization energy 
between the $3d$ and $2p$ states. 
The energy of the $2p$ level relative to the $3d$ levels is determined from 
the charge-transfer energy $\Delta$ defined by 
$\Delta=E^{d}-E^{p}+15U(3d^6)-10U(3d^5)$ with $E^d$ being an average of 
$E^d_m$. Here $U(3d^6)$ and $U(3d^5)$ are the multiplet-averaged $d$-$d$ 
Coulomb interaction in the $3d^6$ and $3d^5$ configurations, 
which are defined by $U=F^{0}-\left(2/63\right)F^{2}-\left(2/63\right)F^{4}$.

\begin{figure}[h]
\begin{center}\includegraphics[%
  scale=0.5]{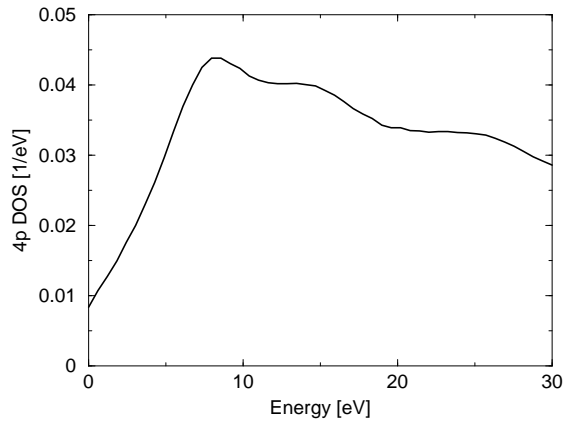}
\caption{Density of states of the $4p$ band. It is constructed from 
the experimental $K$-edge absorption spectra\cite{Kubota2004}
with cutting off the low-energy tail coming from the life-time width 
of the core hole. The high-energy side is arbitrarily cut-off.
The integrated value is normalized to unity.
\label{DOS}}
\end{center}
\end{figure}

The $H^{4p}$ represents the energy of the $4p$ states, 
where $p'_{{\bf k}\eta'\sigma}$ is the annihilation operator of the $4p$ 
state with momentum ${\bf k}$, $\eta'=x,y,z$, and spin $\sigma$.
The $4p$ states form an energy band $\epsilon_{4p}({\bf k})$.
The density of states (DOS) of the $4p$ band is inferred from
the $K$-edge absorption spectra\cite{Kubota2004} as shown in Fig.~\ref{DOS}.
The $H^{4p-2p}_{\rm hyb}$ represents the hybridization between the $4p$
and oxygen $2p$ states, where the annihilation operator of the local $4p$
orbital $p'_{\eta'\sigma}$ may be expressed as 
$p'_{\eta'\sigma}=(1/\sqrt{N_0})\sum_{\bf k} p'_{{\bf k}\eta'\sigma}$
($N_0$ is the discretized number of ${\bf k}$-points).

The hybridization matrices $t^{3d-2p}_{m\eta}(j)$ and 
$t^{4p-2p}_{\eta'\eta}(j)$ are defined for the Fe atom at the off-center 
position. We evaluate these values by modifying the Slater-Koster two-center 
integrals for the Fe atom at the central position of the octahedron 
with the assumption that $(pd\sigma)_{2p,3d}$, $(pd\pi)_{2p,3d}\propto d^{-4}$,
and $(pp\sigma)_{4p,2p}$, $(pp\pi)_{4p,2p}\propto d^{-2}$ for $d$ being 
the Fe-O distance.\cite{Harrison2004}
Table \ref{table.1} lists the parameter values used in this paper,
which are consistent with the values in previous calculations for
Fe$_3$O$_4$.\cite{Chen2004,Igarashi2008}

\begin{table}
\caption{\label{table.1}
Parameter values for a FeO$_{6}$ cluster in the $3d^5$ configuration,
in units of eV. The Slater-Koster two-center integrals are defined
for the Fe atom at the center of the octahedron.}
\begin{ruledtabular}
\begin{tabular}{lrlr}
$F^0(3d,3d)$ & 6.39 & $(pd\sigma)_{2p,3d}$ & -1.9 \\
$F^2(3d,3d)$ & 9.64 & $(pd\pi)_{2p,3d}$ & 0.82 \\
$F^4(3d,3d)$ & 6.03 & $(pp\sigma)_{2p,4p}$ & 3.5 \\
$\zeta_{3d}$ & 0.059 & $(pp\pi)_{2p,4p}$ & -1.0 \\
$\Delta$ & 3.3 &  &  \\

\end{tabular}
\end{ruledtabular}
\end{table} 

\subsection{Ligand field and effective hybridization between $4p$ and 
$3d$ states}

Instead of directly treating $H^{3d-2p}$ and $H^{4p-2p}$,
we introduce the effective Hamiltonian to include the covalency effect.
The ligand field Hamiltonian on the $3d$ states is given by
the second-order perturbation as
\begin{equation}
 \tilde{H}^{3d-3d}= \sum_{mm'\sigma} \tilde{t}_{mm'}^{3d-3d} 
  d_{m\sigma}^{\dagger}d_{m'\sigma} + {\rm H.c.},
\end{equation}
with
\begin{equation}
 \tilde{t}^{3d-3d}_{mm'} = \sum_{j\eta} t^{3d-2p}_{m\eta}(j)
 t^{3d-2p}_{m'\eta}(j)/\Delta,
\end{equation}
where the sum over $j$ is taken on neighboring O sites, and
$\Delta=3.3$ eV is the charge transfer energy defined above.
In addition to the ligand field corresponding to the cubic symmetry, 
we have a field proportional to $\delta^2$, which causes extra splittings 
of $3d$ levels in conformity with the form of Eq.~(\ref{eq.crys}).

The effective hybridization between the $4p$ and $3d$ states is similarly
given as
\begin{equation}
 \tilde{H}^{4p-3d}= \sum_{\eta'm\sigma} \tilde{t}_{\eta'm}^{4p-3d} 
  p'^{\dagger}_{\eta'\sigma}d_{m\sigma} + {\rm H.c.},
\end{equation}
with
\begin{equation}
 \tilde{t}^{4p-3d}_{\eta'm} = \sum_{j\eta} t^{4p-2p}_{\eta'\eta}(j)
       t^{3d-2p}_{m\eta}(j)/(E^{4p}-E^{2p}),
\end{equation}
where $E^{4p}$ is the average of the $4p$-band energy, which
is estimated as $E^{4p}-E^{2p}\approx 17$ eV. 
The coefficient $\tilde{t}^{4p-3d}_{\eta'm}$ is nearly proportional to the 
shift $\delta$ of the Fe atom from the center of the octahedron,
again in conformity with the form of Eq.~(\ref{eq.crys}).

\section{Absorption process on F\lowercase{e}}

The interaction between the electromagnetic wave and electrons is described by
\begin{equation}
 H_{\rm int} = -\frac{1}{c}\int {\bf j}({\bf r})\cdot{\bf A}({\bf r})
 {\rm d}^3{\bf r},
\label{eq.int}
\end{equation}
where ${\bf j}$ represents the current density operator,
and the electromagnetic field ${\bf A}({\bf r})$ for linear polarization
is defined as 
\begin{equation}
 {\bf A}({\bf r}) = \sum_{\bf q}
 \sqrt{\frac{2\pi\hbar c^2}{V\omega_{\bf q}}}
  {\bf e}c_{\bf q}{\rm e}^{i \textbf{q}\cdot\textbf{r}} + {\rm H.c.},
\end{equation}
with $c_{\bf q}$ and ${\bf e}$ 
being the annihilation operator of photon and the unit vector of polarization,
respectively. We approximate this expression into a sum of the contributions 
from each Fe atom:
\begin{equation}
 H_{\rm int} = -\frac{1}{c}\sum_{{\bf q},i}{\bf j}({\bf q},i)\cdot
 {\bf A}({\bf q},i) + {\rm H.c.},
\end{equation}
with
\begin{eqnarray}
 {\bf j}({\bf q},i) &=& \sum_{nn'} 
  \left[\int {\rm e}^{i{\bf q}\cdot({\bf r}-{\bf r}_i)}
  {\bf j}_{nn'}({\bf r}-{\bf r}_i){\rm d}^3({\bf r}-{\bf r}_i)\right]
    a_{n}^{\dagger}(i)a_{n'}(i) , 
\label{eq.localcurrent}\\
 {\bf A}({\bf q},i) &=& \sqrt{\frac{2\pi\hbar c^2}{V\omega_{\bf q}}}
   {\bf e}c_{\bf q}{\rm e}^{i \textbf{q}\cdot\textbf{r}_i},
\end{eqnarray}
where the local current operator may be described by
\begin{eqnarray}
 {\bf j}_{nn'}({\bf r}-{\bf r}_i) &=& \frac{ie\hbar}{2m}
 \big[(\nabla \phi^{*}_n)\phi_{n'}
  - \phi_n^{*}\nabla\phi_{n'}\big] - \frac{e^2}{mc}{\bf A}\phi_n^{*}\phi_{n'}¡
  \nonumber\\
  &+& \frac{e\hbar}{mc}c\nabla\times [\phi_n^{*}{\bf S}\phi_{n'}].
\label{eq.current}
\end{eqnarray}
The integration in Eq.~(\ref{eq.localcurrent}) is carried out around site $i$,
and $a_n(i)$ is the annihilation operator of electron with the local 
orbital with the wave function $\phi_n({\bf r}-{\bf r}_i)$. 
The $e$ and $m$ are the charge and the mass of electron, 
and $\hbar{\bf S}$ is the spin operator of electron.
The second term in Eq.~(\ref{eq.current}), which describes the scattering
of photon, will be neglected in the following discussion. 
The approximation made by taking account of the process only on Fe atoms
may be justified at the core-level spectra, but less accurate in the optical
region. The spectra arising from the magnetoelectric effect, however,
are expected to be described rather well by the present approximation,
since such effects mainly take place on Fe atoms.

\begin{figure}[h]
\begin{center}\includegraphics[%
  scale=0.5]{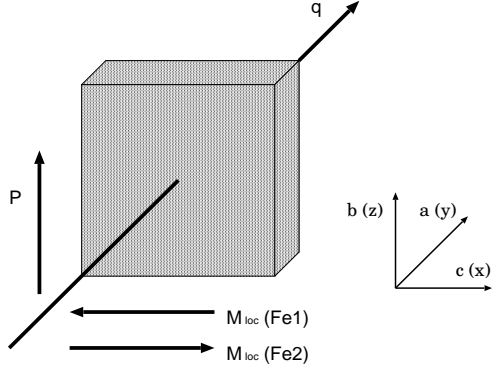}
\caption{Geometry of absorption. Light propagates along the $a$ axis
with polarization along the $b$ axis or the $c$ axis.
The electric dipole moment is along the $b$ axis. The sublattice magnetization 
is directed to the negative direction of the $c$ axis at Fe1 
sites and to the reverse at Fe2 sites, when the external magnetic field 
is applied to the positive direction of the $c$ axis.
When the external magnetic field is reversed, the sublattice magnetization
is reversed.
\label{fig.setup}}
\end{center}
\end{figure}

For later convenience, we write the interaction between the matter and
the photon in a form,
\begin{equation}
 H_{\rm int} = -e\sum_{\bf q}\sqrt{\frac{2\pi}{V\hbar\omega_{\bf q}}}
    \sum_{i} T({\bf q},{\bf e},i)
     c_{\bf q} {\rm e}^{i \textbf{q}\cdot\textbf{r}_i} + {\rm H.c.}.
\end{equation}
To be specific in connection with the experimental set-up,\cite{Jung2004}
we consider the situation that the photon propagates along the $a$-axis 
with linear polarization, as illustrated in Fig.~\ref{fig.setup}.

\subsection{$E1$ transition}

The transition operator $T({\bf q},{\bf e},i)$  for the $E1$ transition 
is given by putting ${\rm e}^{i{\bf q}\cdot({\bf r}-{\bf r}_j)}=1$ in 
Eq.~(\ref{eq.localcurrent}). 
Therefore it is independent of the propagation direction of photon.
For the polarization along the z-axis, the first term in Eq.~(\ref{eq.current})
is rewritten by employing the following relation
\begin{equation}
 \int \phi_n^* \frac{\partial}{\partial z}\phi_{n'} 
{\rm d}^3{\bf r} = -\frac{m}{\hbar^2}(\epsilon_n-\epsilon_{n'})
  \int \phi_n^* z\phi_{n'} {\rm d}^3{\bf r},
\end{equation}
where $\epsilon_n$ and $\epsilon_{n'}$ are energy eigenvalues with $\phi_n$ and 
$\phi_{n'}$, respectively. The $4p$ and $3d$ states are assigned to
$\phi_n$ and $\phi_{n'}$, respectively.
Hence the transition operator $T^{E1}$ is expressed as 
\begin{equation}
 T^{E1}({\bf q},{\bf e},i) =
  iB^{E1} \sum_{i\eta m\sigma} N_{\eta m}^{E1}
     [p'^{\dagger}_{\eta\sigma}(i)d_{m\sigma}(i)
     -d_{m\sigma}^{\dagger}(i)p'_{\eta\sigma}(i)], 
\label{eq.e1mat}
\end{equation}
where $i$ runs over Fe sites. The $N_{\eta m}^{E1}$'s are given by
$N_{x,zx}^{E1}=1/\sqrt{5}$, 
$N_{y,yz}^{E1}=1/\sqrt{5}$, 
$N_{z,3z^2-r^2}^{E1}=2/\sqrt{15}$ for the polarization along the $z$ axis, 
$N_{x,x^2-y^2}^{E1}=1/\sqrt{5}$, 
$N_{x,3z^2-r^2}^{E1}=-1/\sqrt{15}$, 
$N_{y,xy}^{E1}=1/\sqrt{5}$, 
$N_{z,zx}^{E1}=1/\sqrt{5}$ for the polarization along the $x$ axis, and
$N_{x,xy}^{E1}=1/\sqrt{5}$, 
$N_{y,x^2-y^2}^{E1}=-1/\sqrt{5}$, 
$N_{y,3z^2-r^2}^{E1}=1/\sqrt{15}$, 
$N_{z,yz}^{E1}=1/\sqrt{5}$ for  the polarization along the $y$ axis,
respectively.
The coefficient $B^{E1}$ is defined by
\begin{equation}
 B^{E1} = (\epsilon_{4p}-\epsilon_{3d})\int_{0}^{\infty}
   r^3 R_{4p}(r)R_{3d}(r){\rm d}r, 
\end{equation}
where $R_{3d}(r)$, $R_{4p}(r)$ are radial wave-functions 
of $3d$, $4p$ states with energy $\epsilon_{3d}$, $\epsilon_{4p}$ 
in the Fe atom. The energy difference $\epsilon_{4p}-\epsilon_{3d}$ 
is not directly related to the absorbed photon energy. 
Within the HF approximation in the 1s$^2$3d$^5$4p$^{0.001}$-configuration 
of an Fe atom,\cite{Cowan1981} we estimate it as
$B^{E1} \approx 7.7\times 10^{-8}$ cm$\cdot$eV.

\subsection{$E2$ transition}

The transition operator for the $E2$ transition is given from the second term 
in the expansion ${\rm e}^{i{\bf q}\cdot({\bf r}-{\bf r}_i)} \approx 1
+i{\bf q}\cdot({\bf r}-{\bf r}_i)+ \cdots$ in Eq.~(\ref{eq.localcurrent}).
Let the photon be propagating along the $y$-axis with the polarization 
parallel to the $z$-axis. Then we could derive a relation,
\begin{equation}
 \int \phi_n^* y\frac{\partial}{\partial z}\phi_{n'} 
{\rm d}^3{\bf r} = -\frac{m}{\hbar^2}(\epsilon_n-\epsilon_{n'})
  \int \phi_n^* \frac{yz}{2}\phi_{n'} {\rm d}^3{\bf r} 
  +\frac{i}{2}\int \phi_n^* L_x\phi_{n'} {\rm d}^3{\bf r},
\label{eq.e2int}
\end{equation}
where $\hbar L_x$ is the orbital angular momentum operator. 
The last term should be moved into the terms of the $M1$ transition. 
In the first term of Eq.~(\ref{eq.e2int}), the relevant states for $\phi_n$ 
and $\phi_{n'}$ are both $3d$ states, and $\epsilon_n-\epsilon_{n'}$ 
may be an order of the ligand field energy, which is less 
than $1$ eV. 
Since $\langle r^2\rangle$ is estimated within the HF approximation as 
\cite{Cowan1981}
\begin{equation}
 \int_{0}^{\infty}r^4 R_{3d}^2(r){\rm d}r = 3.3\times 10^{-17}{\rm cm}^2,
\end{equation}
we notice that the contribution from the $E2$ transition is smaller than 
that from the $M1$ transition discussed in the next subsection. 

\subsection{$M1$ transition}

From the third term in Eq.~(\ref{eq.current}), we have a relation
\begin{equation}
\int {\rm e}^{i{\bf q}\cdot({\bf r}-{\bf r}_i)} 
 \nabla\times(\phi_n^{*}{\bf S}\phi_{n'}) {\rm d}^3{\bf r} =
 -i{\bf q}\times \int \phi_n^{*}{\bf S}\phi_{n'} 
 {\rm e}^{i{\bf q}\cdot({\bf r}-{\bf r}_i)} {\rm d}^3{\bf r} \approx 
 -i{\bf q}\times \int \phi_n^{*}{\bf S}\phi_{n'} {\rm d}^3{\bf r}.
\end{equation}
Adding the contribution of the last term of Eq.~(\ref{eq.e2int}), we have a
factor ${\bf L}+2{\bf S}$ in the transition operator. 
The $3d$ states are assigned to $\phi_{n}$ and $\phi_{n'}$.
Hence the transition operator for the $M1$ transition is given by
\begin{equation}
 T^{M1}({\bf q},{\bf e},i) = 
   i|\textbf{q}|B^{M1}\sum_{imm'\sigma\sigma'} N^{M1}_{m\sigma,m'\sigma'}
   d^{\dagger}_{m\sigma}(i)d_{m'\sigma'}(i),
\label{eq.m1mat}
\end{equation}
where $B^{M1}=\hbar^2/2m=3.8\times 10^{-16}{\rm cm}^2\cdot{\rm eV}$.
For the photon propagating along the $y$ axis with polarizations
along the $z$ and $x$ axes, we have 
$N^{M1}_{m\sigma,m'\sigma'}=\langle m\sigma|L_x+2S_x|m'\sigma'\rangle$
and $N^{M1}_{m\sigma,m'\sigma'}=\langle m\sigma|-(L_z+2S_z)|m'\sigma'\rangle$,
respectively. 

\section{Calculation of absorption spectra}

Restricting the processes only on Fe atoms, we sum up cross sections 
at Fe sites to obtain the absorption intensity $I(\omega_{\bf q},{\bf e})$.
Dividing it by the incident flux $c/V$, we have
\begin{equation}
 I(\omega_{\bf q},{\bf e}) \propto
  \frac{4\pi^2e^2}{\hbar^2c}\frac{1}{\omega_{\bf q}}
  \sum_i \sum_f |\langle \Psi_f(i)|T({\bf q},{\bf e},i)|\Psi_g(i)\rangle|^2
  \delta(\hbar\omega_{\bf q}+E_g-E_f),
\end{equation}
where $T({\bf q},{\bf e},i)= T^{E1}({\bf q},{\bf e},i) 
+ T^{M1}({\bf q},{\bf e},i)$,
and $|\Psi_g(i)\rangle$ and $|\Psi_f(i)\rangle$ represent the ground and 
the final states with energy $E_g$ and $E_f$ at site $i$, respectively.
The sum over $f$ is taken over all the excited state at Fe sites.

We first calculate the energy eigenstates $|\Phi_n(d^5)\rangle$ 
with eigenenergy $E_n(d^5)$ in the $3d^5$-configuration,
and $|\Phi_n(d^4)\rangle$ with eigenenergy $E_n(d^4)$ in the 
$3d^4$-configuration, by diagonalizing the Hamiltonian 
$H_{3d}+\tilde{H}^{3d-3d}$. As already stated in Sec.~II,
the exchange field ${\bf H}_{\textrm{xc}}$ in Eq.~(\ref{eq.H3d}) is assumed to
be directed to the negative direction of the $c(x)$ axis at Fe1 sites 
and the reverse direction at Fe2 sites when the external magnetic field is
applied to the positive direction of the $c$ axis.
All the directions could be reversed by reversing the external magnetic
field, since the actual compound is a ferrimagnet.
The shift $\delta$ of Fe atoms along the $b$-axis is assumed 
$\delta=0.26 \textrm{\AA}$ 
at Fe1 sites and $\delta=-0.11 \textrm{\AA}$ at Fe2 sites, respectively. 

As regards the lowest energy state $|\Phi_g(d^5)\rangle$,
we have the state $^{6}A_{1}$ under the trigonal crystal field, if we
disregard the exchange field and the spin-orbit interaction. The inclusion of
these interactions could induce the orbital moment $\langle L_x\rangle$, 
but its absolute value is given less than $0.004$. 
Two types of octahedrons give the same angular momentum.

Within the first order perturbation with the effective hybridization 
$\tilde{H}^{4p-3d}$, we could express the ground state $|\Psi_g(i)\rangle$ 
and the optical final states $|\Psi_f(i)\rangle$ as
\begin{eqnarray}
|\Psi_g(i)\rangle &=& |\Phi_g(d^5)\rangle \nonumber \\
               &+& \sum_{n{\bf k}\eta\sigma}
 |\Phi_n(d^4),{\bf k}\eta\sigma\rangle 
  \frac{1}{E_g(d^5)-(E_n(d^4)+\epsilon_{4p}({\bf k}))} \nonumber\\
 &\times&\langle\Phi_n(d^4),{\bf k}\eta\sigma| 
 \tilde{H}^{4p-3d}|\Phi_g(d^5)\rangle, 
\label{eq.initial}\\
|\Psi_f(i)\rangle &=& |\Phi_f(d^5)\rangle \nonumber \\
               &+& \sum_{n{\bf k}\eta\sigma}
  |\Phi_{n}(d^4),{\bf k}\eta\sigma\rangle 
  \frac{1}{E_{f}(d^5)-(E_{n}(d^4)+\epsilon_{4p}({\bf k}))} \nonumber\\ 
  &\times& \langle\Phi_{n}(d^4),{\bf k}\eta\sigma| 
  \tilde{H}^{4p-3d}|\Phi_{f}(d^5)\rangle, 
\label{eq.final.opt}
\end{eqnarray}
with $E_g=E_{g}(d^5)$ and $E_f=E_{f}(d^5)$
being the lowest and excited energies in the $d^5$ configurations,
respectively.
Here $|\Phi_n(d^4),{\bf k}\eta\sigma\rangle$ represents the state of 
four electrons in the 3d states and one electron on 
the $4p$ states specified by $\eta$($=x,y,z$), spin $\sigma$,
and momentum ${\bf k}$. 
The sum over ${\bf k}$ may be replaced by the integral with the $4p$ DOS.
The explicit dependence on site $i$ is abbreviated
in the right hand side of Eqs.~(\ref{eq.initial}) and (\ref{eq.final.opt}).
From these wave-functions we obtain the expressions of optical transition 
amplitudes at site $i$ by
\begin{eqnarray}
 M^{E1}({\bf q},{\bf e},i;f) &\equiv& \langle\Psi_f(i)|
 T^{E1}({\bf q},{\bf e},i)|\Psi_g(i)\rangle 
  \nonumber\\
  &=& \sum_{n{\bf k}\eta\sigma} \langle\Phi_f(d^5)|T^{E1}({\bf q},{\bf e},i)
      |\Phi_n(d^4),{\bf k}\eta\sigma\rangle \nonumber\\
  &\times& 
  \frac{1}{E_g(d^5)-E_n(d^4)-\epsilon_{4p}({\bf k})} 
  \langle\Phi_n(d^4),{\bf k}\eta\sigma|\tilde{H}^{4p-3d}|\Phi_g(d^5)\rangle
 \nonumber\\
  &+& \sum_{n{\bf k}\eta\sigma} \langle\Phi_f(d^5)|\tilde{H}^{4p-3d} 
      |\Phi_n(d^4),{\bf k}\eta\sigma\rangle \nonumber\\
  &\times& 
  \frac{1}{E_f(d^5)-E_n(d^4)-\epsilon_{4p}({\bf k})} 
  \langle\Phi_n(d^4),{\bf k}\eta\sigma|T^{E1}({\bf q},{\bf e},i)
  |\Phi_g(d^5)\rangle, 
\label{eq.ME1}\\
 M^{M1}({\bf q},{\bf e},i;f) &\equiv& \langle\Psi_f(i)|
  T^{M1}({\bf q},{\bf e},i)|\Psi_g(i)\rangle 
  \nonumber\\
  &=& \langle\Phi_f(d^5)|T^{M1}({\bf q},{\bf e},i)|\Phi_g(d^5)\rangle.
\label{eq.MM1}
\end{eqnarray}
With these amplitudes, we have
\begin{equation}
 I(\omega_{\bf q},{\bf e}) \propto 
 \frac{1}{\hbar\omega_{\bf q}} \sum_{i}\sum_{f} 
   |M^{E1}({\bf q},{\bf e},i;f)+M^{M1}({\bf q},{\bf e},i;f)|^2 
  \delta(\hbar\omega_{\bf q}+E_g(d^5)-E_f(d^5)).
\end{equation} 

Now we examine the symmetry relation of the amplitudes.
First, let the propagating direction of photon be reversed with keeping
other conditions. The magnetic field associated with the photon is reversed, 
$N^{M1}$'s in Eq.~(\ref{eq.m1mat}) change their signs. 
Since other conditions are the same, we have the new amplitudes 
$(M^{E1})'=M^{E1}$, $(M^{M1})'=-M^{M1}$.
Second, let the local magnetic moment at each Fe atom be reversed 
with keeping the same shifts from the center of octahedron.
The reversing of the local magnetic moment corresponds to taking 
the complex conjugate of wave functions.
Considering Eq.~(\ref{eq.ME1}) together with Eq.~(\ref{eq.e1mat}),
we have $(M^{E1})'=-(M^{E1})^{*}$. 
Also, considering Eq.~(\ref{eq.MM1}) together with Eq.~(\ref{eq.m1mat}),
we have $(M^{M1})'=(M^{M1})^{*}$. 
Third, let the shifts of Fe atoms from the center of octahedron be reversed 
with keeping the same local magnetic moment, which means the reversal of
the direction of the local \emph{electric} dipole moment. 
This operation gives rise to
reversing the sign of $\tilde{H}^{4p-3d}$ but no change in the $3d$ states 
with the $3d^5$- and $3d^4$-configurations, because the ligand field 
$\tilde{H}^{3d-3d}$ changes according to $\delta^2$. 
As a result, we have the new amplitude $(M^{E1})'=-M^{E1}$ 
from Eq.~(\ref{eq.e1mat}) but no change $(M^{M1})'=M^{M1}$.

As already stated, the direction of the local magnetic moment could be
reversed by reversing the direction of the applied magnetic field, 
since the actual material is a ferrimagnet with slightly deviating from 
a perfect antiferromagnet. 
We define $\Delta I(\omega_{\bf q},{\bf e})$ by the difference between 
the absorption intensity with the applied magnetic field along the 
positive direction of the $c$ axis and that with the field along 
the reverse direction.
From the second symmetry relation mentioned above, we have
\begin{eqnarray}
 \Delta I(\omega_{\bf q},{\bf e}) &\propto& 
 \frac{2}{\hbar\omega_{\bf q}} \sum_{i}\sum_{f} 
   \Bigl{[} \left\{M^{E1}({\bf q},{\bf e},i;f) \right\}^{*}
   M^{M1}({\bf q},{\bf e},i;f) \nonumber\\
  &+& \left\{M^{M1}({\bf q},{\bf e},i;f)\right\}^{*}
     M^{E1}({\bf q},{\bf e},i;f) \Bigr{]}
  \delta(\hbar\omega_{\bf q}+E_g(d^5)-E_f(d^5)).
\label{eq.inte1m1}
\end{eqnarray} 
Considering the sign change, we infer from the above symmetry relations that
\begin{equation}
 \Delta I(\omega_{\bf q},{\bf e}) \propto \frac{{\bf q}}{|\textbf{q}|} \cdot
  \sum_{i} {\bf P}_{\rm loc}(i)\times {\bf M}_{\rm loc}(i) ,
\label{eq.toroidal}
\end{equation}
where ${\bf P}_{\rm loc}(i)$ and ${\bf M}_{\rm loc}(i)$ are the electric 
and the magnetic dipole moment of Fe atom at site $i$, respectively 
(${\bf P}_{\rm loc}(i)\propto \boldsymbol{\delta}_{i}\equiv (0,0,\delta)$).
This relation may be regarded as a lowest order expansion with respect to
$\boldsymbol{\delta}_i$ and ${\bf M}_{\rm loc}(i)$.
The right hand side of Eq.~(\ref{eq.toroidal}) is the sum of the local
toroidal moment $\boldsymbol{\tau}(i)$ 
($\equiv\boldsymbol{\delta}_i\times {\bf M}_{\rm loc}(i)$).
\cite{Popov1998}

\begin{figure}[h]
\begin{center}\includegraphics[%
  scale=0.8]{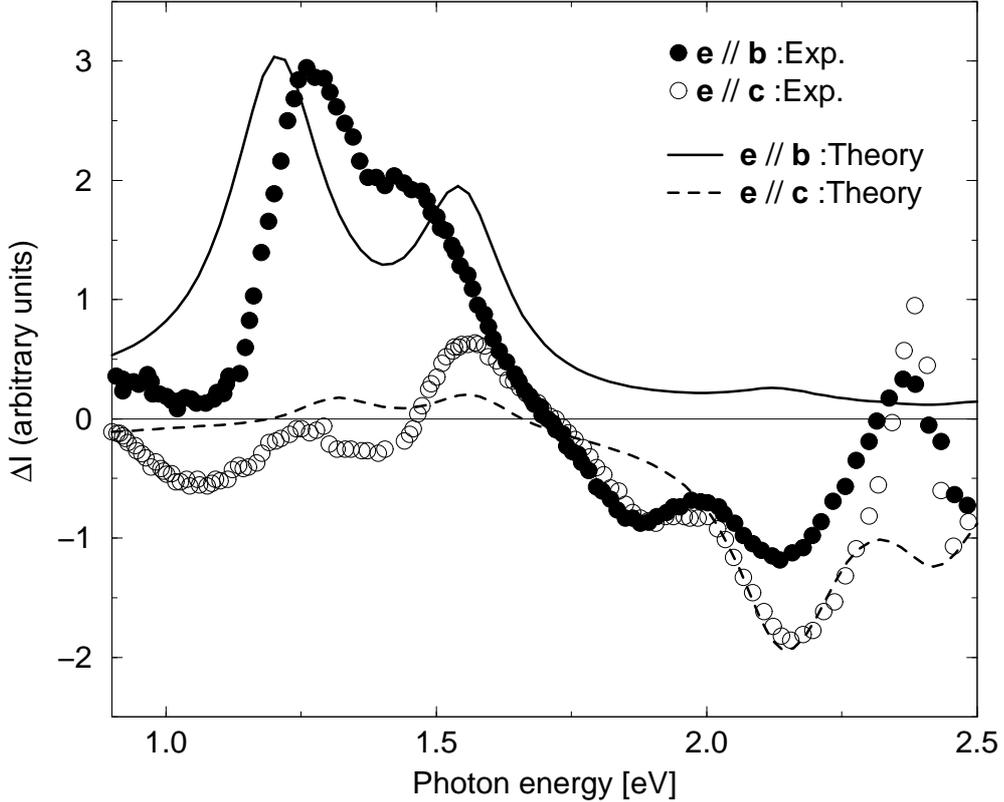}
\caption{Difference of absorption intensities 
$\Delta I(\omega_{\bf q},{\bf e})$ 
as a function of photon energy $\hbar\omega_{\bf q}$
between the applied magnetic field along the positive and negative 
directions of the $c$ axis.
Photon propagates along the positive $a$ axis with polarization vector
${\bf e}$ along the $b$ and $c$ axes, respectively.
Experimental data are taken from Ref.~[\onlinecite{Jung2004}].
\label{fig.magnetoelectric}}
\end{center}
\end{figure}

Figure \ref{fig.magnetoelectric} shows the calculated 
$\Delta I(\omega_{\bf q},{\bf e})$ as a function of $\omega_{\bf q}$, 
in comparison with the experiment.
We have replaced the $\delta$-function $\delta(x)$ in Eq.~(\ref{eq.inte1m1})
by a Lorentzian form $(\gamma/\pi)/(x^2+\gamma^2)$ with $\gamma=0.1$ eV.
The calculated peak height at $\sim 1.2$ eV is set to be the same 
as the experimental one for the polarization ${\bf e}$ along the $b$ axis. 
We have a two-peak structure around 
$\hbar\omega_{\bf q}=1.0-1.5$ eV in agreement with the experiment,
but could not reproduce a dip found experimentally around 
$\hbar\omega_{\bf q}=1.7-2.3$ eV.
On the other hand, without further adjustment,
we have a considerable dip around $\hbar\omega_{\bf q}=2.0-2.5$ eV 
for ${\bf e}$ along the $c$ axis,
in agreement with the experiments.

Fe atoms are under the cubic symmetry without displacement,
and the lowest and low-lying excited states are characterized as
$^{6}A_{1}$, $^{2}T_{2}$, $^{4}T_{1}$, $^{4}T_{2}$ with neglecting 
the spin-orbit interaction and the exchange field.\cite{Tanabe1954}
The excitation energies for $^{2}T_{2}$, $^{4}T_{1}$, and $^{4}T_{2}$ 
are estimated $1.34$, $1.59$, $2.45$ eV, respectively, 
within the present cluster model. Note that the direct absorption processes
$^{6}A_{1}\to {^{2}T_{2}}$, $^{6}A_{1}\to {^{4}T_{1}}$, and 
$^{6}A_{1}\to {^{4}T_{2}}$ are forbidden. 
The displacement of the Fe atom generates a trigonal field and makes the 
energy levels of the excited states split. The spin-orbit interaction 
and the exchange field further modify these states. 
The magnetoelectric spectra around $1.0-1.5$ eV and around $1.7-2.3$ eV 
might be interpreted as transitions to the states dispersed from 
$^{2}T_{2}$ and $^{4}T_{1}$, and those from $^{4}T_{2}$, respectively.

\section{Concluding Remarks}

We have studied the magnetoelectric effects on the optical absorption spectra 
in a polar ferrimagnet GaFeO$_3$. 
We have considered the $E1$, $E2$, and $M1$ processes on Fe atoms, 
and have performed a microscopic calculation of the magnetoelectric spectra 
using a cluster model of FeO$_6$.
The cluster consists of an octahedron of O atoms and an Fe atom displaced
from the center of octahedron. 
We have  disregarded additional small distortions of the octrahedron.
Due to the noncentrosymmetric environment on the Fe atom,
we have an effective  hybridization between the $4p$ and $3d$ states 
through the O $2p$ states and thereby the mixing of the $3d^44p$-configuration 
to the $3d^5$-configuration. 
This mixing makes the $E1$-$M1$ interference process survive and gives rise 
to the magnetoelectric spectra.
We have evaluated the $E1$-$M1$ process by using the energy eigenstates 
given in the $3d^44p$-configuration and the $3d^5$-configuration.
The Coulomb interaction between $3d$ electrons and the hybridization 
are assumed to be nearly the same as previous cluster calculations.
\cite{Chen2004,Igarashi2008}
We have obtained the magnetoelectric spectra as a function of photon energy
in the optical region $1.0-2.5$ eV, in agreement with the experiment.

In the experiment, the conventional absorption spectra, a part independent 
of the direction of magnetization, were measured with intensity about three 
orders of magnitude larger than the magnetoelectric part.\cite{Jung2004}
On the other hand, in the present approach considering only the local process 
on Fe atoms, the ``total" intensity, which is given by the $E1$-$E1$ and 
$M1$-$M1$ processes, is estimated as merely one order of magnitude larger 
than that of the $E1$-$M1$ process. This suggests that other processes 
such as the transition from the valence band to the conduction band 
involving Ga and O atoms may add larger contributions.
As far as the magnetoelectric spectra are concerned, however,
the present approach considering only the local process on Fe atoms
is expected to work well, since the $E1$-$M1$ interference process
could take place only on Fe atoms.
Finally, from a different point of view, 
we would like to comment that the approach of considering the 
multiple scattering of a $4p$ electron in the noncentrosymmetric potential 
and the Coulomb interaction between the $4pd^4$ and the $d^5$ configurations
may improve the above situation. The critical study is left in future.

We have concentrated on the spectra in the optical region.
In the x-ray region, the magnetoelectric effects have also been studied.
\cite{Goulon2002,Matteo2002,Carra2003,Kubota2004,Lovesey2007}
Since the core electron is excited there, the local approach 
in this paper would be better applicable to the x-ray region than 
to the optical region,
where the $E1$-$E2$ (not $E1$-$M1$) interference process gives rise to
the magnetoelectric spectra.
It may be interesting to analyze microscopically the nonreciprocal 
directional dichroism observed in the Fe pre-$K$-edge x-ray absorption 
in GaFeO$_3$\cite{Kubota2004} by using a similar cluster model.
In this context, we would like to comment that
the magnetoelectric effect on the resonant x-ray scattering spectra 
has been analyzed at the Fe pre-$K$-edge in Fe$_3$O$_4$,\cite{Igarashi2008}
where Fe atoms at $A$ sites are located at the center of tetrahedron 
in noncentrosymmetric environment.

\begin{acknowledgments}

This work was partly supported by Grant-in-Aid 
for Scientific Research from the Ministry of Education, Culture, Sport, 
Science, and Technology, Japan.

\end{acknowledgments}

\bibliography{paper}

\end{document}